\documentclass[10pt,prl,twocolumn,showpacs,aps,floats,floatfix,superscriptaddress]{revtex4}

\usepackage{epsfig}
\begin{document}

\title{Optimal Traffic Networks}

\author{Marc Barth\'elemy}
\affiliation{School of Informatics and Biocomplexity Center,
Indiana University,Eigenmann Hall, 1900 East Tenth Street,
Bloomington IN 47406}
\affiliation{CEA-Centre d'Etudes de
Bruy{\`e}res-le-Ch{\^a}tel, D\'epartement de Physique Th\'eorique et
Appliqu\'ee BP12, 91680 Bruy\`eres-Le-Ch\^atel, France}

\author{Alessandro Flammini}
\affiliation{School of Informatics and Biocomplexity Center,
Indiana University,Eigenmann Hall, 1900 East Tenth Street,
Bloomington IN 47406}

\date{\today} \widetext

\begin{abstract}

Inspired by studies on the airports' network and the physical
Internet, we propose a general model of weighted networks via an
optimization principle. The topology of the optimal network turns out
to be a spanning tree that minimizes a combination of topological and
metric quantities.  It is characterized by a strongly heterogeneous
traffic, non-trivial correlations between distance and traffic and a
broadly distributed centrality. A clear spatial hierarchical
organization, with local hubs distributing traffic in smaller regions,
emerges as a result of the optimization. Varying the parameters of the
cost function, different classes of trees are recovered, including in
particular the minimum spanning tree and the shortest path tree. These
results suggest that a variational approach represents an alternative
and possibly very meaningful path to the study of the structure of
complex weighted networks.

\end{abstract}


\maketitle



Transportation and communication infrastructures such as the airports'
network and the physical Internet are characterized by broad
distributions of traffic~\cite{Barrat:2004,Barrat:2005}, betweenness
centrality~\cite{Goh:2002} and in some case also of
degree~\cite{Amaral:2005}. Strong, non-linear traffic-distance and
traffic-connectivity correlations have also been
reported~\cite{Barrat:2004,Barrat:2005}. Modelization attempts with
ingredients such as random weights, dynamical rules or weight-topology
coupled dynamics (see eg.~\cite{Barrat:2005} and refs. therein) have
mainly focused on growth processes and variational approaches have
instead mainly been used only in practical problems by road traffic
engineers~\cite{Wardrop:1952}. Both the problem of optimal
traffic~\cite{ahuja} on a network and of optimal
networks~\cite{Jungnickel:1999} have a long tradition in mathematics
and physics. It is well known, for example that the laws that describe
the flow of currents in a resistor network~\cite{kirchoff} can be
derived by minimizing the energy dissipated by the
network~\cite{snell}. On the other hand, optimal networks have shown
to be relevant in the study of mammalians circulatory
system~\cite{mahon}, food webs~\cite{Garlas:2003}, general
transportation networks~\cite{Banavar:1999}, metabolic
rates~\cite{West:1997}, river networks~\cite{colaiori}, and gas
pipelines or train tracks~\cite{Gastner:2004}. All these studies share
the fact that the nodes of the network are embedded in a d-dimensional
euclidean space which implies that the degree is almost always limited
and the connections restricted to `neighbours' only.  A second broad
class of optimal networks where spatial constraints are absent has
been also recently investigated. It has been shown, for example, that
optimization of both the average shortest path and the total length
can lead to small-world networks~\cite{Mathias:2001}, and more
generally, degree correlations~\cite{Berg:2002} or scale-free
features~\cite{Valverde:2002} can emerge from an optimization
process. Cancho and Sole~\cite{Cancho:2001} showed that the
minimization of the average shortest path and the link density leads
to a variety of networks including exponential-like graphs and
scale-free networks. Guimera {\it et al.}~\cite{Guimera:2002} studied
networks with minimal search cost and found two classes of networks:
star-like and homogeneous networks. Finally, Colizza {\it et
  al.}~\cite{Colizza:2004} studied networks with the shortest route
and the smallest congestion and showed that this interplay could lead
to a variety of networks when the number of links per node is changed.

In this paper mostly inspired by studies on airports'
networks~\cite{Amaral:2005,Barrat:2004}, we investigate the important
case for which nodes are embedded in a 2-dimensional plane but links
are not constrained (as for air routes) to connect `neighbours'. We
propose a cost function that depends both on the length and the
traffic carried by the links and show that the resulting optimal
network is hierarchically organized in space and displays a complex
traffic structure. We consider a set of $N$ points (`airports')
randomly distributed in a square of unitary area and would like to
build a network (air routes) that connects all points. The cost or
weigth $w_{ij}$ associated to `traveling' along a link $(i,j)$ is a
function of both the length $d_{ij}$ of the link and of the traffic
$t_{ij}$ it carries. In the air-network analogy, the quantity $t_{ij}$
represents the number of passengers on the link $(i,j)$ and is symmetric
$t_{ij}=t_{ji}$. To travel from a generic node $i_0$ to another
generic node $i_p$ along a specific path $\{
i_0,i_1,i_2,\dots,i_{p-1},i_p \}$ the cost to be paid is the sum of
the weights $w_{k,k+1}$ associated to the links that compose the path,
and when more than one path is available we assume that the most
economical one is chosen
\begin{equation}
{\cal C} _{i_{0}, i_{p}} = \min _{p \in {\cal P} (i_{0},i_{p} ) } \sum_{e \in p} w_{e}
\end{equation}
where the minimisation is over all paths $p$ belonging in the set of
paths ${\cal P}(i_0,i_p)$ going from $i_0$ to $i_p$ ($w_e$ is the
weight of the edge $e$). If there is no path between a pair of nodes
the corresponding cost will be taken as equal to infinity. This choice
ensures that the optimal network will be connected. The global
quantity ${\cal E}_0$ we wish to minimize is then the average cost to
pay to travel from a generic node to another
\begin{equation}
{\cal E}_0 (\{t_{ij}\}) =\frac{2}{N(N-1)}\sum_{i<j}{\cal C}_{i,j}
\label{eq:energy}
\end{equation}
Our purpose is therefore to find the traffic $\{ t^{*}_{ij} \}$
carried by the links and which minimises~(\ref{eq:energy}), with the
only constraints that all $t_{ij} \ge 0$ and that the total traffic
$T=\sum_{i<j}t_{ij}$ is fixed. In this paper, we choose as the weight
of a link $e$ the ratio of its length to traffic: $w_e=d_e/t_e$ (with
this choice, the value of $T$ fixes the scale of traffic and does not
affect the topology of the optimal network since a rescaling of the
total energy by a constant factor will not affect the
minimization~\cite{Note}). Although this choice is not the most
general, it naturally verifies the expectation that the weight
increases with $d_{ij}$ and decreases with $t_{ij}$. This last
condition can be easily understood in the case of transportation
networks and means that it is more economic to travel on links with a
large traffic, reducing the effective distance---or the marginal
cost---of the connection. We search for the minimum-realizing traffic
using a zero temperature metropolis algorithm. The elementary move
consists in transferring a random fraction of the traffic carried by a
link to another one (the total traffic being fixed). We choose at
random two links $(i,j)$ and $(i',j')$ and transfer weights between
them according to
\begin{eqnarray}
t_{ij}&\to& t_{ij}-\alpha t_{ij}\\
t_{i'j'}&\to& t_{i'j'}+\alpha t_{ij}
\end{eqnarray}
where $\alpha$ is a uniform random number between zero and one. The
sign of $\alpha$ is positive with probability $p$ and negative with
probability $1-p$ (if one of the links has a zero weight, the transfer
can only be made in one direction, in other cases $p=0.5$ ensures a
quick convergence). If after the transfer the weight of a link is
zero, the corresponding link is deleted. The minimum-cost path between
two points is recalculated at any step with Djikstra's
algorithm~\cite{Jungnickel:1999}. We compute the energy difference
$\Delta={\cal E}_0'-{\cal E}_0$ and only if it is negative the
transfer is accepted. We test a number of order ${\cal O}(N^2)$ of
such transfers which converges to an optimal network which minimizes
the energy ${\cal E}_0$. The initial topology is a complete graph with
random weights on the links. As we show below the optimal solution is
characterized by a non trivial topology and spatial organization which
results as the compromise of two opposing forces: the need for short
routes and the traffic concentration on as few paths as possible. The
interplay between topology and traffic naturally induces the observed
correlations between degree, distance and traffic itself.


Numerical simulations shows that the optimal network is a tree. A
simple example supporting this finding is obtained by considering an isocele
triangle $ABC$ with $d(A,C)=d(B,C)=d$ and $d(A,B)=d'$, optimization
leads to the values $t_{AC}=t_{BC}\approx T/2$ and $t_{AB}\approx 0$
when $d\gg d'$. The minimum energy is thus (at leading order in $d$)
${\cal E}_0\approx d/t_{AC}+d/t_{BC}\approx 4d/T$. When we remove the
link BC and thus kill the loop, the traffic on AC becomes
approximately twice higher, ie. $t_{AC}\approx T$ (and $t_{BC}\approx
t_{AB}\approx 0$) but the minimum energy at leading order is ${\cal
  E}'_0\approx 2d/t_{AC}\approx 2d/T$ which is lower than ${\cal
  E}$. This example shows that optimization reduces the number of
links joining nodes in the same regions and increases the traffic on
the remaining links. Loops between nodes in the same neighborhood
become then redundant.
\begin{figure}
\centerline{
\epsfysize=0.64\columnwidth{\epsfbox{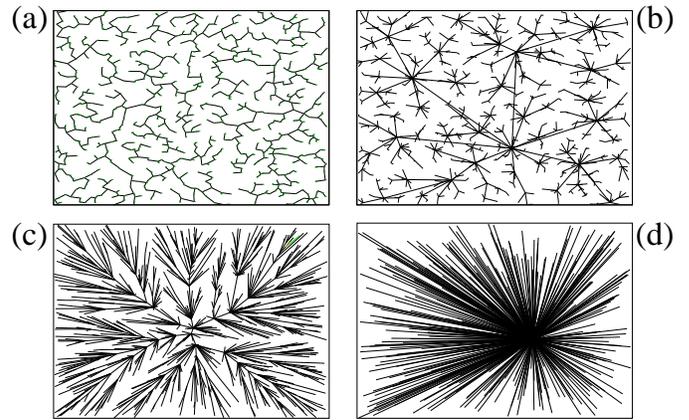}}
}
\caption{ Different spanning trees obtained for different values of
$(\mu,\nu)$ in Eq.~(\ref{eq:munu}) obtained for the same set of $N=1000$
nodes. (a) Minimum spanning tree obtained for $(\mu,\nu)=(0,1)$. In
this case the total distance is minimized. (b) Optimal traffic tree
obtained for $(\mu,\nu)=(1/2,1/2)$. In this case we have an interplay
between centralization and minimum distance resulting in local
hubs. (c) Minimum euclidean distance tree obtained for
$(\mu,\nu)=(1,1)$. In this case centrality dominates over distance and a
`star' structure emerges with a few dominant hubs. (d) Optimal
betweenneess centrality tree obtained for $(\mu,\nu)=(1,0)$. In this
case we obtain the shortest path tree which has one star hub (for the
sake of clarity, we omitted some links in this last figure).}
\label{fig:examples}
\end{figure}

The optimal network being a tree enormously simplifies the computation
of the energy. Since only a single path exists between any two nodes
in a tree, the energy (\ref{eq:energy}) can be rewritten as ${\cal
  E}_0=\sum_{e\in{\cal T}}b_e\frac{d_e}{t_e}$ where $b_e$ is the
edge-betweenness~\cite{Freeman:1977} and counts the number of times
that $e$ belongs to the shortest path between two nodes. The optimal
traffic (with the same constraints as above) is given by
$t_e=T\sqrt{b_ed_e}/\sum_e\sqrt{b_ed_e}$ and the topology of the
`optimal traffic tree' (OTT) can then be obtained by minimizing
\begin{equation}
{\cal E}=\sum_{e\in{\cal T}}\sqrt{b_ed_e}
\label{eq:bd}
\end{equation}
The minimal configuration can now be searched by rewiring links.
Replacing link $(i,j)$ by $(i,j')$ modifies only the centralities along
the path between $j$ and $j'$ which implies that our calculation has a
complexity of order ${\cal O}(N)$ and allows computations over very
large networks (the same algorithm was used in the context of river
networks~\cite{sandro}). We expect to obtain something very different
from the classical (Euclidean-) minimum spanning
tree (MST)~\cite{Jungnickel:1999} since ${\cal E}$ involves a combination of
metric (the distance) and topological (the betweenness)
quantities. The expression Eq.~(\ref{eq:bd}) suggests an interesting
generalization given by the optimization of
\begin{equation}
{\cal E}_{\mu\nu}=\sum_{e\in{\cal T}} b_{e}^{\mu}d_{e}^{\nu}
\label{eq:munu}
\end{equation}
where $\mu$ and $\nu$ control the relative importance of distance
against topology as measured by centrality. Fig.~(\ref{fig:examples})
shows examples of spanning trees obtained for different values of
$(\mu,\nu)$. For $(\mu,\nu)=(0,1)$ one obtains the Euclidean Minimum
Spanning Tree [Fig.~(\ref{fig:examples}a)] which can also be
obtained by minimizing the total weight $\sum_e w_e$ and gives a
traffic $t_{e}=T\sqrt{d_{e}}/\sum \sqrt{d_{e}}$.  For
$(\mu,\nu)=(1/2,1/2)$ we obtain the OTT [Fig.~(\ref{fig:examples}b)]
which displays an interesting interplay between distance and shortest
path minimization (see below).  For $(\mu,\nu)=(1,1)$, the energy is
proportional to the average shortest weighted path (with weights equal
to euclidean distance [Fig.~(\ref{fig:examples}c)].  When
$(\mu,\nu)=(1,0)$, the energy (\ref{eq:munu}) is proportional to the
average betweenness centrality and therefore to the average shortest
path $\sum_e b_e\propto \ell$.  The tree $(1,0)$ shown in
Fig.~(\ref{fig:examples}d) is thus the shortest path tree (SPT) with
an arbitrary ``star-like'' hub (a small non zero value of $\nu$ would
select as the star the closest node to the gravity center).  The
minimization of Eq.~(\ref{eq:munu}) thus provides a natural
interpolation between the MST and the SPT, a problem which was
addressed in previous studies~\cite{Khuller:1995}. The degree
distribution for all cases considered above [with the possible
exception $(\mu,\nu)=(1,1)$---a complete inspection of the plane
$(\mu,\nu)$ is left for future studies] is not broad, possibly as a
consequence of spatial constraints. In particular, the degree
distribution for the OTT is well fitted by an exponential function.

\begin{figure}
\centerline{ \epsfysize=0.55\columnwidth{\epsfbox{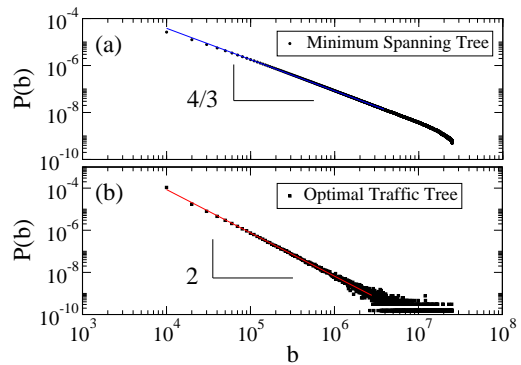}} }
\caption{ Betweenness centrality distribution for the MST and for the
OTT. The lines are power law fits and give for the MST the
theoretical result $\delta=4/3$ and for the OTT the value
$\delta\simeq 2.0$ ($N=10^4$, $100$ configurations).}
\label{fig:bc}
\end{figure}

It has been shown that trees can be classified in `universality
classes'~\cite{takayasu,dhar} according to the size distribution of
the two parts in which a tree can be divided by removing a link (or
the sub-basins areas distribution in the language of river
network). We define $A_i$ and $A_j$ as the sizes of the two parts in
which a generic tree is divided by removing the link $(i,j)$. The
betweeness $b_{ij}$ of link $(i,j)$ can be written as
$b_{ij}=\frac{1}{2} [ A_i (N-A_i) + A_j (N-A_j) ]$, and the
distributions of $A's$ and $b's$ can be easily derived one from the
other. It is therefore not surprising that the same exponent $\delta$
characterizes both $P(A)\sim A^{-\delta}$ and $P(b)$. While we obtain
the value $\delta=4/3$ for the MST~\cite{takayasu}, for the OTT we
obtain (Fig.~\ref{fig:bc}) an exponent $\delta\simeq 2$, a value also
obtained for trees grown with preferential attachment
mechanism~\cite{Albert:2001} (see also~\cite{delos} for a supporting
argument). Interestingly, most real-world networks are also described
by this value $\delta\simeq 2$~\cite{Goh:2002}. The OTT thus tends to
have a more uniform centrality with respect to the
MST~\cite{Shlomo:2005}, with important consequences on the
vulnerability of the network since there is no clearly designated
`Achille's heel' for the OTT.

\begin{figure}
\centerline{
\includegraphics*[angle=+90,width=0.30\textwidth]{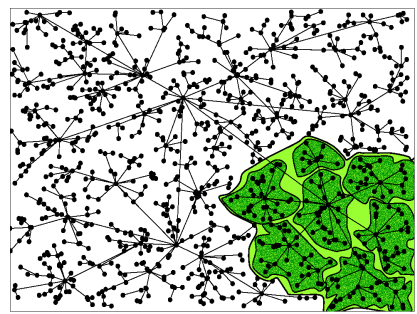}
}
\caption{Hierarchical organization emerging for the optimal traffic
tree $(\mu,\nu)=(1/2,1/2)$ ($N=1000$ nodes). Longer links lead to
regional hubs which in turn connect to smaller hubs distributing
traffic in smaller regions.}
\label{fig:regions}
\end{figure}
\begin{figure}
\centerline{\epsfysize=0.38\columnwidth{\epsfbox{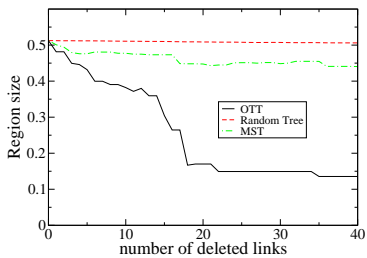}}}
\caption{ Average euclidean size of the largest cluster remaining
  after deleting links ranked according to their length (in decreasing
  order). This plot is obtained for one typical configuration of size
  $N=1000$ and a square of area set to one. The decrease observed for
  the OTT is consistent with a hierarchical spatial organization as it
  is visually evident from Fig.~\ref{fig:regions}.}
\label{fig:ave_d}
\end{figure}

We now investigate the traffic properties of the OTT and we find that
the traffic is scaling as $t_{ij}\sim d_{ij}^{\tau}$ with $\tau\approx
1.5$ showing that large traffic is carried over large distance and is
then dispatched on smaller hubs that distribute it on still smaller
regions. Despite the limited range of degrees, we also observe for the
strength~\cite{Barrat:2004} $s_i=\sum_j t_{ij}$ a superlinear behavior
with the degree. This result demonstrates the existence of
degree-traffic correlations as observed for the airport network for
example~\cite{Barrat:2004} can emerge from a global optimization
process. The spatial properties of the OTT are also remarkable and
displays (Fig.~\ref{fig:regions}) a hierarchical spatial organization
where long links connect regional hubs, that, in turn are connected to
sub-regional hubs, etc. This hierarchical structure can be probed by
measuring the average euclidean distance between nodes belonging to
the largest cluster obtained by deleting recursively the longest
link. For the OTT (Fig.~\ref{fig:ave_d}), we observe a decrease of the
region size, demonstrating that longer links connect smaller regions,
a feature absent in non-hierarchical networks such as the MST, the SPT
or the random tree (Fig.~\ref{fig:ave_d}).


In summary, we showed that the emergence of complex structure in
traffic organization could be explained by an optimization
principle. In particular, strong correlations between distance and
traffic arise naturally as a consequence of optimizing the average
weighted shortest path. In the optimal network, long-range links carry
large traffic and connect regional hubs dispatching traffic on a
smaller scale ensuring an efficient global distribution. These results
suggest that the organization of the traffic on complex networks and
more generally architecture of weighted networks could in part result
from an evolutionary process. The optimal networks obtained here are
trees, but some transportation networks contain loops which reflect
the fact that other ingredients are needed in order to describe
them. Our results however suggest that some transportation networks
could possibly be seen as the superposition of many trees. It would be
interesting to check if this is the case for the world-wide airport
network which results from the superposition of individual airline
company networks which are probably close to trees. Finally, this
study led us to propose a generalization of the usual minimum spanning
tree by introducing the centrality and allows one to interpolate from
the MST to the shortest path tree. This generalization however
deserves further studies and raises interesting question such as the
crossover from different tree universality classes.


Acknowledgments.\--- We thank Vittoria Colizza and Alessandro Vespignani for
interesting discussions and suggestions.




\begin{thebibliography}{99}
\bibitem{Note} We chose $t_{ij}$ to be integer for faster simulations
  but the results are the same with real $t_{ij}$.
\end{thebibliography}
\end{document}